\documentclass[a4paper,12pt]{article}
\usepackage{amsmath}
\usepackage{amssymb}
\usepackage{latexsym}
\usepackage{enumerate}
\newcommand{\be}{\begin{eqnarray}}
\newcommand{\ee}{\end{eqnarray}}

\newcommand{\bsub}{\begin{subequations}}
\newcommand{\esub}{\end{subequations}}

\newcommand{\disfrac}[1][2]{\displaystyle\frac}
\begin{document}
\title{{\bf Towards Canonical Quantum Gravity for Geometries
Admitting Maximally Symmetric Two-dimensional Surfaces}}
\vspace{1cm}
\author{\textbf{T. Christodoulakis}\thanks{tchris@phys.uoa.gr}\,, \textbf{G. Doulis}\thanks{gdoulis@phys.uoa.gr}\,,
\textbf{Petros A. Terzis}\thanks{pterzis@phys.uoa.gr}\\
{\it Nuclear and Particle Physics Section, Physics Department,}\\{\it University of Athens, GR 157--71 Athens}\\
\textbf{E. Melas}\thanks{evangelosmelas@yahoo.co.uk}\\
{\it Technological Educational Institution of Lamia}\\
{\it Electrical Engineering Department, GR 35--100, Lamia }\\
\textbf{Th. Grammenos}\thanks{thgramme@uth.gr}\\
{\it Department of Mechanical Engineering, University of Thessaly,}\\
{\it GR 383--34 Volos}\\
\textbf{G.O. Papadopoulos}\thanks{gopapad@mathstat.dal.ca}\\
{\it Department of Mathematics and Statistics, Dalhousie University}\\
{\it Halifax, Nova Scotia, Canada B3H 3J5}\\
\textbf{A. Spanou}\thanks{aspanou@central.ntua.gr}\\
{\it School of Applied Mathematics and Physical Sciences},\\
{\it National Technical University of Athens, GR 157--80, Athens}}
\date{}
\maketitle
\begin{center}
\textit{}
\end{center}
\vspace{-1cm}
\numberwithin{equation}{section}
\newpage
\begin{abstract}
The 3+1 (canonical) decomposition of all geometries admitting
two-dimensional space-like surfaces is exhibited. A proposal
consisting of a specific re-normalization {\bf Assumption} and an
accompanying {\bf Requirement} is put forward, which enables the
canonical quantization of these geometries. The resulting
Wheeler-deWitt equation is based on a re-normalized manifold
parameterized by three smooth scalar functionals. The entire space
of solutions to this equation is analytically given, exploiting the
freedom left by the imposition of the {\bf Requirement} and
contained in the third functional.
\end{abstract}

\section{Introduction}

Dirac's seminal work on his formalism for a self-contained treatment
of systems with constraints \cite{Dirac1}, \cite{Dirac2},
\cite{Dirac3}, \cite{Dirac} has paved the way for a systematic
treatment of constrained systems. Some of the landmarks in the study
of constrained systems have been the connection between constraints
and invariances \cite{Bergman}, the extension of the formalism to
describe fields with half-integer spin through the algebra of
Grassmann variables \cite{Berezin} and the introduction of the BRST
formalism \cite{Becchi}. All the classical results obtained so far
have made up an armoury prerequisite for the quantization of gauge
theories and there are several excellent reviews studying constraint
systems with a finite number of degrees of freedom \cite{Sudarshan}
or constraint field theories \cite{Hanson}, as well as more general
presentations \cite{Sundermeyer}, \cite{Gitman}, \cite{Govaers},
\cite{Henneaux}, \cite{Wipf}. In particular, the conventional
canonical analysis approach of quantum gravity has been initiated by
P.A.M. Dirac \cite{Dirac4}, P.G. Bergmann \cite{Bergmann} and B.S.
deWitt \cite{DeWitt}. For a modern account see \cite{Thiemann}.
\\
In the absence of a full theory of quantum gravity, it is reasonably
important to address the quantization of (classes of) simplified
geometries. The most  elegant way to achieve a degree of
simplification is to impose some symmetry. For example, the
assumption of a $G_3$ symmetry group acting simply transitively on
the surfaces of simultaneity, i.e. the existence of three
independent space-like Killing Vector Fields (KVF), leads to
classical and subsequently quantum homogeneous cosmology (see e.g.
\cite{Ryan}, \cite{chris1}). The imposition of lesser symmetry, e.g.
fewer KVF's, results in the various inhomogeneous cosmologies
\cite{Krasin}. The canonical analysis under the assumption of
spherical symmetry, which is a $G_3$ acting multiply transitively on
two-dimensional space-like subsurfaces of the three-slices, has been
first considered in \cite{Thomi}, \cite{Hajic}. Quantum black holes
have been treated, for instance, in \cite{Zan} while in
\cite{Kief1}, \cite{Kief} a lattice regularization has been employed
to deal with the infinities arising due to the ill-defined nature of
the quantum operator constraints. In this work we consider the
quantization of all geometries admitting two-dimensional surfaces of
maximal symmetry, i.e. spheres (constant positive curvature), planes
(zero curvature) and Gauss-Bolyai-Lobachevsky (henceforth GBL)
spaces (constant negative curvature). In the second section we give
the reduced metrics, the space of classical solutions and the
Hamiltonian formulation of the reduced Einstein-Hilbert action
principle, resulting in one (quadratic) Hamiltonian and one (linear)
momentum first class  constraint. In the third section we consider
the quantization of this constraint system following Dirac's
proposal of implementing the quantum operator constraints as
conditions annihilating the wave-function \cite{Dirac}. Our
guide-line is a conceptual generalization of the quantization scheme
developed in \cite{Kuchar1}, \cite{Kuchar2} for the case of
constraint systems with finite degrees of freedom, to the present
case, which is enabled through the use of a specific
re-normalization {\bf Assumption} and an associated {\bf
Requirement}. After the symmetry reduction, the system still
represents an one-dimensional field theory since all remaining
metric components depend on time and the radial coordinate.
Nevertheless, we manage to extract a Wheeler-deWitt equation in
terms of three smooth scalar functionals of the reduced geometries.
The exploitation of a residual freedom left by the imposition of the
{\bf Requirement} enables us to acquire the entire solution space to
this equation. Finally, some concluding remarks are included in the
discussion, while the proof of the existence for the third
functional is given in the Appendix.

\section{Possible Metrics, Classical Solutions and Hamiltonian Formulation}

Our starting point is the two-dimensional spaces of positive, zero
and negative constant curvature. Their line elements are
respectively:
\begin{equation}\label{lineelem}
ds^2=d\theta^2+\sin^2\theta\,d\phi^2,\qquad
ds^2=d\theta^2+\theta^2\,d\phi^2,\qquad
ds^2=d\theta^2+\sinh^2\theta\,d\phi^2
\end{equation}
with an obvious range of the coordinates for each case. The corresponding (maximal)
symmetry groups are generated by the following KVF's:
\begin{alignat}{9}
\xi_1 & = \frac{\partial}{\partial\phi}, &\quad
\xi_2 & =-\cos\phi\frac{\partial}{\partial\theta}+\cot\theta\sin\phi\frac{\partial}{\partial\phi}, &\quad
\xi_3 & =\sin\phi\frac{\partial}{\partial\theta}+\cot\theta\cos\phi\frac{\partial}{\partial\phi}\label{killspe}\\
&&&&& \nonumber\\
\xi_1 & =\frac{\partial}{\partial\phi}, &\quad
\xi_2 & =-\cos\phi\frac{\partial}{\partial\theta}+\frac{\sin\phi}{\theta}\frac{\partial}{\partial\phi},&\quad
\xi_3 & =\sin\phi\frac{\partial}{\partial\theta}+\frac{\cos\phi}{\theta}\frac{\partial}{\partial\phi}\label{killpla}\\
&&&&& \nonumber\\
\xi_1 & =\frac{\partial}{\partial\phi}, &\quad
\xi_2 & =-\cos\phi\frac{\partial}{\partial\theta}+\coth\theta\sin\phi\frac{\partial}{\partial\phi}, &\quad
\xi_3 & =\sin\phi\frac{\partial}{\partial\theta}+\coth\theta\cos\phi\frac{\partial}{\partial\phi}\label{killhyp}
\end{alignat}
We next promote these KVF's to four-dimensional fields by adding to
each and every of them the zero-sum
$0\,\frac{\partial}{\partial\,t}\,+0\,\frac{\partial}{\partial\,r}$.
We then enforce these vector fields as symmetries of a generic
space-time metric $g_{IJ}(t,r,\theta,\phi)$, i.e. we demand that its
Lie derivative with respect to all three fields for each family
vanishes. We thus arrive at three classes of metrics, collectively
described by the following line element:
\begin{equation}\label{metrics}
\begin{split}
ds^2=& \left(-\alpha(t,r)^2+\frac{\beta(t,r)^2}{\gamma(t,r)^2}\right)\,dt^2
+2\,\beta(t,r)\,d\,t\,d\,r+\gamma(t,r)^2\,d\,r^2 \\
& +\psi(t,r)^2\,d\,\theta^2+\psi(t,r)^2\,f(\theta)^2\,d\,\phi^2
\end{split}
\end{equation}
where $f(\theta)=\sin\theta$ in the case of spherical symmetry, $f(\theta)=\theta$ in the case of plane symmetry,
and $f(\theta)=\sinh\theta$ in the case of the GBL symmetry.

In order to attain the classical space of solution for these
geometries one can, exploiting the freedom to change coordinates in
the $(t,r)$ subspace, bring the upper left block of the metric in
conformally flat form and readily solve the vacuum Einstein's field
equations. The result is given, in the light-cone coordinates
$u=\displaystyle{\frac{t+r}{2}},\,v=\displaystyle{\frac{-t+r}{2}}$,
by the following line elements:
\begin{equation}\label{metrics1}
ds^2=2\,\epsilon\,\frac{A+2\,\psi(u,v)}{4\,\psi(u,v)}\,du\,dv
+\psi(u,v)^2\,d\theta^2+\psi(u,v)^2\,f(\theta)^2\,d\phi^2,
\end{equation}
where
\[\psi(u,v)= -\frac{A}{2}\left(1+\mbox{ProductLog}(-\frac
{\exp^{-\frac{\lambda(A+u+\epsilon\,v)}{A}}}{A})\right),\]
$\epsilon=\pm 1$ and $\lambda=\pm 1$ for $f(\theta)=\sin\theta$,
$\lambda=\pm i$ for $f(\theta)=\sinh\theta$; $\mbox{ProductLog}(z)$
is the principal solution for $w$ to the equation $z=w\,\exp^w$. The
corresponding to the plane symmetric case $f(\theta)=\theta$
line-element is given by:
\begin{equation}\label{metrics2}
ds^2=2\,\epsilon\,\frac{1}{\psi(u,v)}du\,dv+\psi(u,v)^2 d\theta^2+\psi(u,v)^2\theta^2d\phi^2,
\end{equation}
where
\[\psi(u,v)=
\pm\sqrt{2u+2\epsilon\,v}.\]
The Hamiltonian formulation of the
system (\ref{metrics}) proceeds, briefly, as follows (see, e.g.,
chapter 9 of \cite{Sundermeyer}):\\
We first define the vectors
\[\eta^I=\frac{1}{\alpha(t,r)}(1,-\frac{\beta(t,r)}{\gamma(t,r)^2},0,0),\qquad
F^I=\eta^J_{;J}\,\eta^I-\eta^I_{;J}\,\eta^J,\] where $I$, $J$ are
space-time indices and ``\,$;$\," stands for covariant
differentiation with respect to (\ref{metrics}). Then, utilizing the
Gauss-Codazzi equation, we eliminate all second time-derivatives
from the Einstein-Hilbert action and arrive at an action quadratic
in the velocities, $I=\int d^4x\sqrt{-g}(R-2\Lambda-2\,F^I_{;I})$.
The application of the Dirac algorithm results firstly in the
primary constraints $P_\alpha\equiv\frac{\delta
L}{\delta\dot{\alpha}}\approx0,\,P^\beta\equiv\frac{\delta
L}{\delta\dot{\beta}}\approx0$ and the Hamiltonian
\begin{equation}\label{Hamiltonian}
H=\int \left(N^o\,\mathcal{H}_o+N^i\,\mathcal{H}_i\right)dr,
\end{equation}
where
\[N^o=\alpha(t,r),\quad N^1=\frac{\beta(t,r)}{\gamma(t,r)^2}, \quad N^2=0, \quad N^3=0\]
and $\mathcal{H}_o$, $\mathcal{H}_i$ are given by
\bsub\label{constraints}
\begin{eqnarray}\label{Ho}
\mathcal{H}_o &=& \frac{1}{2}\,G^{\alpha\beta}\,\pi_\alpha\,\pi_\beta+V,\\
\label{H1}\mathcal{H}_1 &=& -\gamma\,\pi^{\prime}_\gamma+
\psi^{\prime}\,\pi_\psi, \qquad \mathcal{H}_2=0, \qquad\mathcal{H}_3=0,
\end{eqnarray}
\esub while the indices $\{\alpha,\beta\}$ take the values
$\{\gamma,\psi\}$ and $^{\prime}=\displaystyle{\frac{d}{d\,r}}$. The
reduced Wheeler-deWitt super-metric $G^{\alpha\beta}$ reads
\begin{equation}\label{supermetric}
G^{\alpha\beta}=
\begin{pmatrix}
\displaystyle{\frac{\gamma}{4\,\psi^2}} & -\displaystyle{\frac{1}{4\,\psi}} \cr
& \cr
-\displaystyle{\frac{1}{4\,\psi}}
& 0
\end{pmatrix},
\end{equation}
while the potential $V$ is
\begin{equation}\label{potential}
V=-2\,\epsilon\,\gamma+2\Lambda\gamma\,\psi^2-2\frac{\psi^{\prime\, 2}}{\gamma}+4\left(\frac{\psi\,\psi^{\prime}}{\gamma}\right)^{\prime}
\end{equation}
with $\epsilon=\{1,0,-1\}$ for the families (\ref{metrics}) of
two-dimensional subspaces with positive, zero or negative constant
curvature, respectively. The requirement for preservation, in time, of
the primary constraints leads to the secondary constraints
\begin{equation}\label{conweak}
\mathcal{H}_o\approx0,\qquad\qquad \mathcal{H}_1\approx0
\end{equation}
At this stage, a tedious but straightforward calculation produces
the following open Poisson bracket algebra of these constraints:
\be\label{alge}
\{\mathcal{H}_o(r),\mathcal{H}_o(\tilde{r})\}&=&\left[\frac{1}{\gamma^2(r)}\mathcal{H}_1(r)+
\frac{1}{\gamma^2(\tilde{r})}\mathcal{H}_1(\tilde{r})\right]\delta^{\prime}(r,\tilde{r}),\nonumber\\
&&\nonumber\\
\{\mathcal{H}_1(r),\mathcal{H}_o(\tilde{r})\}&=&\mathcal{H}_o(r)\delta^{\prime}(r,\tilde{r}),\\
&&\nonumber\\
\{\mathcal{H}_1(r),\mathcal{H}_1(\tilde{r})\}&=&\mathcal{H}_1(r)
\delta^{\prime}(r,\tilde{r})-\mathcal{H}_1(\tilde{r})\delta(r,\tilde{r})^{\prime},\nonumber
\ee indicating that they are first class and also signaling the
termination of the algorithm. Thus, our system is described by
(\ref{conweak}) and the dynamical Hamilton-Jacobi equations
$\disfrac{d\,\pi_\gamma}{d\,t}=\{\pi_\gamma,H\}$,
$\disfrac{d\,\pi_\psi}{d\,t}=\{\pi_\psi,H\}$. One can readily check
(as one must always do with reduced action principles) that these
four equations, when expressed in the velocity phase-space with the
help of the definitions $\disfrac{d\,\gamma}{d\,t}=\{\gamma,H\}$,
$\disfrac{d\,\psi}{d\,t}=\{\psi,H\}$, are completely equivalent to
the four independent Einstein's field equations satisfied by
(\ref{metrics}).

We end up this section by noting a few facts concerning the
transformation properties of $\gamma(t,r),\, \psi(t,r)$ and their
derivatives under changes of the radial variable $r$ of the form
$r\rightarrow\tilde{r}=h(r)$. As it can easily be inferred from
(\ref{metrics}):
\begin{equation}\label{tensor}
\tilde{\gamma}(\tilde{r})=\gamma(r)\,\frac{d\,r}{d\,\tilde{r}}, \qquad
\tilde{\psi}(\tilde{r})=\psi(r), \qquad
\frac{d\,\tilde{\psi}(\tilde{r})}{d\,\tilde{r}}=\frac{d\,\psi(r)}{d\,r}\,\frac{d\,r}{d\,\tilde{r}},
\end{equation}
where the $t$-dependence has been omitted for the sake of brevity.
Thus, under the above coordinate transformations, $\psi$ is a scalar,
while $\gamma,\,\psi^{\prime}$ are covariant rank 1 tensors (one-forms), or,
equivalently in one dimension, scalar densities of weight $-1$.
Therefore, the scalar derivative is not $\disfrac{d}{d\,r}$ but rather
$\disfrac{d}{\gamma\,d\,r}$. Finally, if we consider an infinitesimal
transformation $r\rightarrow\tilde{r}=r-\eta(r)$, it is easily seen
that the corresponding changes induced on the basic fields
are:
\begin{equation}\label{lieder}
\delta\,\gamma(r)=(\gamma(r)\,\eta(r))^{\prime}, \qquad\qquad \delta\,\psi(r)=\psi^{\prime}(r)\,\eta(r)
\end{equation}
i.e., nothing but the one-dimensional analogue of the appropriate Lie
derivatives.\\
With the use of (\ref{lieder}), we can reveal the
nature of the action of $\mathcal{H}_1$ on the basic configuration
space variables as that of the generator of spatial diffeomorphisms:
\begin{equation}\label{sdif}
\begin{split}
& \left\{\gamma(r)\,,\,\int d\tilde{r}\,\eta(\tilde{r})\,\mathcal{H}_1(\tilde{r})\right\}=(\gamma(r)\,\eta(r))^{\prime},\\
& \left\{\psi(r)\,,\,\int
d\tilde{r}\,\eta(\tilde{r})\,\mathcal{H}_1(\tilde{r})\right\}=\psi^{\prime}(r)\,\eta(r).
\end{split}
\end{equation}
Thus, we are justified to consider $\mathcal{H}_1$ as the
representative, in phase-space, of an arbitrary infinitesimal
re-parametrization of the radial coordinate. In the same manner we
can also see that the action of the quadratic constraint
$\mathcal{H}_o$ on the basic configuration space variables is
identical to an infinitesimal change of the time coordinate, see
e.g. p. 21 in \cite{Carlip}. The qualitative difference in the
corresponding proof  is that the dynamical equations giving the time
derivatives of the momenta are explicitly used; hence the terms
generator of ``time deformations'' or ``dynamical evolution'' used
for $\mathcal{H}_o$.

Thus, the linear and quadratic constraints are seen to be the
generators of space-time diffeomorphisms, i.e. they represent in
phase space the local ``gauge'' coordinate transformations;
therefore the imposition of their proper quantum analogues will
guarantee the invariance of the ensuing quantum theory under the
aforementioned ``gauge''. Our study is limited, at the present
state, to this goal and is thus not concerned with global issues
like boundary terms.

\section{Quantization}
We are now interested in attempting to quantize this Hamiltonian
system following Dirac's general spirit of realizing the classical
first class constraints (\ref{conweak}) as quantum operator
constraint conditions annihilating the wave function. The main
motivation behind such an approach is the justified desire to
construct a quantum theory manifestly invariant under the ``gauge"
generated by the constraints. To begin with, let us first note that,
despite the simplification brought by the imposition of the symmetry
(\ref{killspe}), (\ref{killpla}), (\ref{killhyp}), the system  is
still a field theory in the sense that all configuration variables
and canonical conjugate momenta depend not only on time ( as is the
case in homogeneous cosmology), but also on the radial coordinate $r$.
Thus, to canonically quantize the system in the Schr\"{o}dinger
representation, we first realize the classical momenta as functional
derivatives with respect to their corresponding conjugate fields
\[\hat{\pi}_{\gamma}(r)=-i\,\disfrac{\delta}{\delta\,\gamma(r)}, \qquad\hat{\pi}_{\psi}(r)=-i\,
\disfrac{\delta}{\delta\,\psi(r)}.\]
We next have to decide on the initial space of state vectors. To
elucidate our choice, let us consider the action of a momentum operator on
some function of the configuration field variables, say
\[\hat{\pi}_{\gamma}(r) \gamma(\tilde{r})^2=
-2i\gamma(\tilde{r}) \delta(\tilde{r},r).\]
The Dirac delta-function
renders the outcome of this action a distribution rather than a
function. Also, if the momentum operator were to act at the same
point as the function, i.e. if  $\tilde{r}=r$, then its action would
produce a $\delta(0)$ and would therefore be ill-defined. Both of
these unwanted features are rectified, as far as expressions linear
in momentum operators are concerned, if we choose as our initial
collection of states all $\emph{smooth functionals}$ (i.e.,
integrals over $r$) of the configuration variables
$\gamma(r),\psi(r)$ and their derivatives of any order. Indeed, as
we infer from the previous example,
\[\hat{\pi}_{\gamma}(r)\int d\tilde{r} \gamma(\tilde{r})^2=
-2i\,\int d\tilde{r}\gamma(\tilde{r})
\delta(\tilde{r},r)=-2i\gamma(r);\]
thus the action of the
momentum operators on all such states will be well-defined (no
$\delta(0)$'s) and will also produce only local functions and not
distributions. However, even so, $\delta(0)$'s will appear as soon
as local expressions quadratic in  momenta are considered, e.g.,
\[\hat{\pi}_{\gamma}(r)\,\hat{\pi}_{\gamma}(r)\int d\tilde{r}
\gamma(\tilde{r})^2=\hat{\pi}_{\gamma}(r) (-2i\int
d\tilde{r}\gamma(\tilde{r})
\delta(\tilde{r},r))=\hat{\pi}_{\gamma}(r)(-2i\gamma(r))=-2
\delta(0).\] An other problem of equal, if not greater, importance
has to do with the number of derivatives (with respect to $r$)
considered: A momentum operator acting on a smooth functional of
degree $\emph{n}$ in derivatives of $\gamma(r),\psi(r)$ will, in
general, produce a function of degree $2\emph{n}$, e.g.,
\[\hat{\pi}_{\gamma}(r)\int d\tilde{r}
\gamma''(\tilde{r})^2=-2i\int d\tilde{r}
\gamma''(\tilde{r})\delta''(\tilde{r},r)=-2i\gamma^{(4)}(r).\] Thus,
clearly, more and more derivatives must be included if we desire the
action of momentum operators to keep us inside the space of
integrands corresponding to the initial collection of smooth
functionals; eventually, we have to consider $n\rightarrow\infty$.
This, in a sense, can be considered as the reflection to the
canonical approach, of the non-re-normalizability results existing
in the so-called covariant approach. Loosely speaking, the way to
deal with these problems is to regularize (i.e., render finite) the
infinite distribution limits, and re-normalize the theory by,
somehow, enforcing $\emph{n}$ to terminate at some finite value.

In the following, we are going to present a quantization scheme of
our system which:
\begin{enumerate}[(a)]
\item  avoids the occurrence of $\delta(0)$'s
\item  reveals the value $\emph{n}=1$ as the only \emph{natural} (i.e. without \emph{ad-hoc} cut-offs) possibility to obtain a
closed space of state vectors
\item  extracts a finite-dimensional Wheeler-deWitt equation governing the quantum
dynamics.
\end{enumerate}
The scheme closely parallels, conceptually, the quantization
developed in \cite{Kuchar1}, \cite{Kuchar2} for finite systems with
one quadratic and a number of linear first class constraints.
Therefore, we deem it appropriate, and instructive, to present a
brief account of the essentials of this construction.

To this end, let us consider a system described by a Hamiltonian of
the form
\be
H & \equiv & \mu X+\mu^i \chi_i \nonumber\\
& = &\mu\,\left(\frac{1}{2}G^{AB}(Q^\Gamma)P_AP_B\,+U^A(Q^\Gamma) P_A+V(Q^\Gamma)\right)\,
+\mu^i\,\phi_i^A(Q^\Gamma)P_A,
\ee
where $A,B,\Gamma\ldots=1,2\ldots, M$
count the configuration space variables and\break $i=1,2,\ldots, N<(M-1)$ numbers
the super-momenta constraints $\chi_i\approx0$, which along with
the super-Hamiltonian constraint $X\approx0$ are assumed to be
first class:
\begin{equation}\label{algekuch}
\{X,X\}=0, \qquad \{X,\chi_i\}=X C_i+C^j_i \chi_j, \qquad
\{\chi_i,\chi_j\}=C^k_{ij} \chi_k,
\end{equation}
where the first (trivial) Poisson bracket has been included only to emphasize
the difference from (\ref{alge}).

The physical state of the system is
unaffected by the ``gauge" transformations generated by $(X,\,\chi_i)$,
{\bf but also} under the following three changes:
\begin{enumerate}[(I)]
\item Mixing of the super-momenta with a non-singular matrix
\[\bar{\chi}_i=\lambda^j_i(Q^\Gamma)  \chi_j\]
\item Gauging of the  super-Hamiltonian with the super-momenta
\[\bar{X}=X+\kappa^{(Ai}(Q^\Gamma)\phi_i^{B)}(Q^\Gamma)P_AP_B+\sigma^i(Q^\Gamma)\phi_i^A(Q^\Gamma)P_A\]
\item Scaling of the super-Hamiltonian
\[\bar{X}=\tau^2(Q^\Gamma)X.\]
\end{enumerate}
Therefore, the geometrical structures on the configuration space
that can be inferred from the super-Hamiltonian are really
equivalence classes under actions (I), (II) and (III); for example
(II), (III) imply that the super-metric $G^{AB}$ is known only up to
conformal scalings and additions of the super-momenta coefficients
$\bar{G}^{AB}=\tau^2(G^{AB}+\kappa^{(Ai}\phi_i^{B)})$. It is thus
mandatory that, when we Dirac-quantize the system, we realize the
quantum operator constraint conditions on the wave-function in such
a way as to secure that the whole scheme is independent of actions
(I), (II), (III). This is achieved by the following steps:
\begin{enumerate}[(1)]
\item Realize the linear operator constraint conditions with the momentum
operators to the right
\[\hat{\chi}_i\Psi=0\leftrightarrow\phi_i^A(Q^\Gamma)\frac{\partial\,\Psi(Q^\Gamma)}{\partial\,Q^A}=0,\]
which maintains the geometrical meaning of the linear constraints
and produces the $M-N$ independent solutions to the above equation
$q^\alpha(Q^\Gamma),\,\alpha=1,2,\ldots, M-N$ called physical variables,
since they are invariant under the transformations generated by the
$\hat{\chi}_i$'s.
\item In order to make the final states physical with respect to the
``gauge" generated by the quadratic constraint $\hat{X}$ as well:

Define the induced structure $g^{\alpha\beta}\equiv
G^{AB}\disfrac{\partial\,q^\alpha}{\partial\,Q^A}\disfrac{\partial\,q^\beta}{\partial\,Q^B}$
and realize the quadratic in momenta part of $X$ as the conformal
Laplace-Beltrami operator based on $g_{\alpha\beta}$. Note that in
order for this construction to be self consistent, all components of
$g_{\alpha\beta}$ must be functions of the physical coordinates
$q^\gamma$. This can be proven to be so by virtue of the classical
algebra the constraints satisfy (for specific quantum cosmology
examples see \cite{chris1}).
\end{enumerate}

We are now ready to proceed with the quantization of our system, in
close analogy to the scheme above outlined. In order to realize the
equivalent to step 1, we first define the quantum analogue of
$\mathcal{H}_1(r)\approx 0$ as
\begin{equation}\label{qulincon}
\hat{\mathcal{H}}_1(r)\Phi=0\leftrightarrow
-\gamma(r)\,(\frac{\delta\,\Phi}{\delta\,\gamma(r)})'+\psi'(r)\,\frac{\delta\,\Phi}{\delta\,\psi(r)}=0.
\end{equation}
As explained in the beginning of the section, the action of
$\hat{\mathcal{H}}_1(r)$ on all smooth functionals is well defined,
i.e., produces no $\delta(0)$'s. It can be proven that, in order for
such a functional to be annihilated by this linear quantum operator,
it must be scalar, i.e. have the form
\bsub
\begin{eqnarray}
\label{Phi1}\Phi&=&\int\gamma(\tilde{r})\,L\left(\Psi^{(0)},\Psi^{(1)},\ldots,\Psi^{(n)}\right)d\tilde{r}\\
\label{Psi}\Psi^{(0)}&\equiv&\psi(\tilde{r}),\quad\Psi^{(1)}\equiv\frac{\psi'(\tilde{r})}{\gamma(\tilde{r})},
\ldots,\Psi^{(n)}\equiv\frac{1}{\gamma(\tilde{r})}\frac{d}{d\tilde{r}}
\left(\underset{n-1}{\underbrace{\ldots}}\,\,\psi(\tilde{r})\right)
\end{eqnarray}
\esub where $L$ is any function of its arguments. We note that, as
it is discussed at the end of the previous section,
$\frac{\psi'}{\gamma}$ is the only scalar first derivative of
$\psi$, and likewise for the higher derivatives. The proof of this
statement is analogous to the corresponding result concerning full
gravity \cite{Tracy}: consider an infinitesimal re-parametrization
of $r$ $\tilde{r}=r-\eta(r)$. Under such a change, the left-hand
side of (\ref{Phi1}), being a number, must remain unaltered. If we
calculate the change induced on the right-hand side we arrive at
\begin{equation}
0=\int\left[L\delta\gamma+\gamma\,\frac{\delta L}{\delta\gamma}\delta\gamma+
\gamma\,\frac{\delta L}{\delta\psi}\delta\psi\right]dr=\int[\gamma\,\hat{\mathcal{H}}_1(L)]\eta(r)dr,
\end{equation}
where use of (\ref{lieder}) and a partial integration has been
made. Since this must hold for any $\eta(r)$, the result sought for
is obtained.

We now try to realize step 2 of the programm previously outlined. We
have to define the equivalent of Kucha\v{r}'s induced metric on the
so far space of ``physical'' states $\Phi$ described by (\ref{Phi1})
which are the analogues, in our case, of Kucha\v{r}'s physical
variables $q^\alpha$. Let us start our investigation by considering
\textbf{one} initial candidate of the above form. Then, generalizing
the partial to functional derivatives, the induced metric will be
given by
\begin{equation}\label{physical}
g^{\Phi\Phi}=G^{\alpha\beta}\,\frac{\delta\Phi}{\delta\,x^\alpha}\,
\frac{\delta\Phi}{\delta\,x^\beta}, \quad\mbox{where}\quad x^\alpha=\{\gamma,\psi\}
\end{equation}
and $G^{\alpha\beta}$ is given by (\ref{supermetric}). Note that
this metric is well defined since it contains only first functional
derivatives of the state vectors, as opposed to any second order
functional derivative operator that might be considered as a quantum
analogue of the kinetic part of $\mathcal{H}_o$. Nevertheless,
$g^{\Phi\Phi}$ \emph{is} a local function and not a smooth
functional. It is thus clear that, if we want the induced metric
$g^{\Phi\Phi}$ to be composed out of the ``physical'' states
annihilated by $\hat{\mathcal{H}}_1$, we must establish a
correspondence between local functions and smooth functionals. A way
to achieve this is to adopt the following ansatz: \\
\\
\textbf{Assumption:} \emph{We assume that, as part of the
re-normalization procedure, we are permitted to map local functions
to their corresponding smeared expressions e.g.,
$\psi(r)\leftrightarrow \int d\tilde{r}\psi(\tilde{r})$.}\\
\\
Let us be more specific, concerning the meaning of the above
Assumption. Let $\mathcal{F}$ be the space which contains all local
functions, and define the equivalence relations
\begin{equation}\thicksim: \{f_1(r)\thicksim
f_2(\tilde{r}),\tilde{r}=g(r) \}, \quad \thickapprox:
\{h_1(r)\thickapprox
h_2(\tilde{r})\,\frac{d\,\tilde{r}}{d\,r},\tilde{r}=g(r)
\}\end{equation} for scalars and densities respectively.

Now let $\mathcal{F}_o=\{f\in \mathcal{F}, \mod
(\thicksim,\thickapprox)\}$ and $\mathcal{F_I}$ the space of the
smeared functionals. We define the one to one maps $\mathfrak{G}$,
$\mathfrak{G}^{-1}$
\begin{equation} \mathfrak{G}: \mathcal{F}_o \mapsto
\mathcal{F}_I: \quad \psi(r) \mapsto \int\psi(\tilde{r})\,d\tilde{r}, \quad
\mathfrak{G}^{-1}: \mathcal{F}_I \mapsto \mathcal{F}_o: \quad
\int\psi(\tilde{r})\,d\tilde{r} \mapsto \psi(r)
\end{equation}
The necessity to define the maps $\mathfrak{G}, \mathfrak{G}^{-1}$
on the equivalence classes and not on the individual functions,
stems out of the fact that we are trying to develop a quantum theory
of the geometries \eqref{metrics} and not of their coordinate
representations. If we had tried to define the map $\mathfrak{G}$
from the original space $\mathcal{F}$ to $\mathcal{F}_I$ we would
end up with states which would not be invariant under spatial
coordinate transformations ($\mathbf{r}$-reparameterizations).
Indeed, one can make a correspondence between
local functions and smeared expressions, but smeared expressions
\emph{must} contain another arbitrary smearing function, say $s(r)$.
Then the map between functions and smeared expressions is one to one
(as is also the above map) and is given by multiplying by $s(r)$ and
integrating over $r$; while the inverse map is given by varying
w.r.t. $s(r)$. However, this would be in the opposite direction from
that which led us to the states \eqref{Phi1} by imposition of the
linear operator constraint. As an example consider the action of
this operator on one particular case of the states \eqref{Phi1},
containing the structure $s(r)$ :
\begin{equation}
\hat{\mathcal{H}}_1(r) \int
s(\tilde{r})\,\gamma(\tilde{r})\,\psi(\tilde{r})\,d\tilde{r}=-s'(r)\,\gamma(r)\,\psi(r)
\neq 0 \quad \text{for arbitrary}\, s(r)
\end{equation}
Thus, every foreign to the geometry structure $s(r)$ is not allowed to enter the
physical states.

Now, after the correspondence has been established, we can come to
the basic property the induced metric must have. In the case of
finite degrees of freedom the induced metric depends, up to a
conformal scaling, on the physical coordinates $q^\alpha$ by virtue
of (\ref{algekuch}). In our case, due to the dependence of the
configuration variables on the radial coordinate $r$, the above
property is not automatically satisfied; e.g. the functional
derivative $\frac{\delta}{\delta\psi(r)}$ acting on
$\Psi^{(\emph{n})}$ will produce, upon partial integration of the
$n^{th}$ derivative of the Dirac delta function, a term proportional
to $\Psi^{(2\emph{n})}$. Therefore, since $L$ in \eqref{Phi1}
contains derivatives of $\psi(r)$ up to $\Psi^{(\emph{n})}$, the
above mentioned property must be \emph{enforced}. The need for this
can also be traced to the substantially different first Poisson
bracket in (\ref{alge}), which signals a non trivial mixing between
the dynamical evolution
generator $\mathcal{H}_o$ and the linear generator $\mathcal{H}_1$.\\
Thus, according to the above reasoning, in order to proceed with the
generalization of Kucha\v{r}'s method, we have to demand that:\\
\\
\textbf{Requirement:}
\emph{$L\left(\Psi^{(0)},\ldots,\Psi^{(\emph{n})}\right)$ must be
such that $g^{\Phi\Phi}$ becomes a general function, say
$F\left(\gamma(r)\,L(\Psi^{(0)},\ldots,\Psi^{(\emph{n})})\right)$ of
the integrand of $\Phi$, so that it can be considered a function of
this state: $g^{\Phi\Phi}\overset{Assumption}{\equiv}F\left(
\int\gamma(\tilde{r})\,L(\Psi^{(0)},\ldots,\Psi^{(\emph{n})})d\tilde{r}\right)=F(\Phi)$}.\\
\\
At this point, we must emphasize that the application of the
\textbf{Requirement} in the subsequent development of our quantum
theory will result in very severe restrictions on the form of
\eqref{Phi1}. Essentially, all higher derivatives of $\psi(r)$ (i.e
$\Psi^{(2)}\ldots\Psi^{(\emph{n})})$) are eliminated from $\Phi$
(see \eqref{Phi2}, \eqref{y1,y2,y3} bellow). This might, at first
sight, strike as odd; indeed, the common belief is that all the
derivatives of the configuration variables should enter the physical
states. However, before the imposition of both the linear \emph{and}
the quadratic constrains there are no truly physical states. Thus,
no physical states are lost by the imposition of the
\textbf{Requirement}; ultimately the only true physical states are
the solutions to \eqref{wdw}. Of course, if one insists, one can
keep higher derivatives, say $k$, in the functional. But then, in
order to enforce the \textbf{Requirement} one would have to
eliminate by hand all derivatives higher than $k$. However, to our
view, such an action would seem very \emph{un-natural}. In addition,
although we do not have a concrete rigorous proof, we believe that
this \emph{ad-hoc} elimination would$-$in the present
approach$-$break the space-time covariance of the quantum theory we
are trying to built.

Having clarified the way in which we view the \textbf{Assumption}
and \textbf{Requirement} above, we now proceed to the restrictions
implied by their use.

Let us now turn to the degree of derivatives $(\emph{n})$ of
$\psi(r)$. As we argued before, the functional derivatives
$\frac{\delta}{\delta\psi(r)}$ and $\frac{\delta}{\delta\gamma(r)}$
acting on a functional containing in its integrand
$\Psi^{(\emph{n})}$ will, upon partial integration of the
$\emph{n}^{th}$ derivative of the Dirac delta function, produce a
term proportional to $\Psi^{(2\emph{n})}$ and $\Psi^{(2\emph{n}-1)}$
respectively. More precisely
\begin{eqnarray*}
g^{\Phi\Phi}=\ldots+2\,G^{12}\frac{\delta \Phi}{\delta
\gamma(r)}\frac{\delta\Phi} {\delta\psi(r)}\,.
\end{eqnarray*}
Where the functional derivatives are:
\begin{eqnarray*}
\frac{\delta\Phi}{\delta\psi}&=&\ldots+\int \gamma\,\frac{\partial
L}{\partial \Psi^{(\emph{n})}} \frac{\delta
\Psi^{(\emph{n})}}{\delta \psi}\,d\tilde{r}=\ldots+\int
\gamma\,\frac{\partial L} {\partial
\Psi^{(\emph{n})}}\,\frac{1}{\gamma}\,\frac{d}{d\tilde{r}}\left(\underset{\emph{n}-1}
{\underbrace{\ldots}}\,\,\delta(r,\tilde{r})\right)d\tilde{r}=\\
&=&\ldots-\int \frac{d}{d\tilde{r}}\left(\frac{\partial L}{\partial
\Psi^{(\emph{n})}}\right)
\frac{1}{\gamma}\,\frac{d}{d\tilde{r}}\left(\underset{\emph{n}-2}{\underbrace{\ldots}}
\,\,\delta(r,\tilde{r})\right)d\tilde{r}=\\
&=&\ldots-\int \gamma\,\frac{\partial^2 L}{\partial
\left(\Psi^{(\emph{n})}\right)^2}\,\Psi^{(\emph{n}+1)}
\,\frac{1}{\gamma}\,\frac{d}{d\tilde{r}}\left(\underset{\emph{n}-2}{\underbrace{\ldots}}\,\,
\delta(r,\tilde{r})\right)d\tilde{r}=\\
&& \qquad\qquad\qquad\quad\qquad\vdots\\
&=&\ldots+(-1)^n\int \gamma(\tilde{r})\,\frac{\partial^2 L}{\partial
\left(\Psi^{(\emph{n})}\right)^2}
\,\,\Psi^{(2\emph{n})}\,\delta(r,\tilde{r})\,d\tilde{r}=\\
&=&\ldots+(-1)^\emph{n}\gamma\,\frac{\partial^2 L}{\partial
\left(\Psi^{(\emph{n})}\right)^2}\,\,\Psi^{(2\emph{n})}
\end{eqnarray*}
and
\begin{eqnarray*}
\frac{\delta\Phi}{\delta\gamma}&=&\ldots+\int \gamma\,\frac{\partial
L}{\partial \Psi^{(\emph{n})}} \frac{\delta
\Psi^{(\emph{n})}}{\delta \gamma}\,d\tilde{r}=\ldots+\int
\gamma\,\frac{\partial L} {\partial
\Psi^{(\emph{n})}}\,\frac{1}{\gamma}\,\frac{d}{d\tilde{r}}\left(\underset{\emph{n}-2}
{\underbrace{\ldots}}\,-\frac{\delta(r,\tilde{r})}{\gamma(\tilde{r})^2}\,\psi'(\tilde{r})\right)d\tilde{r}=\\
&=&\ldots+\int \gamma\,\frac{\partial L}{\partial
\Psi^{(\emph{n})}}\,\frac{1}{\gamma}\,\frac{d}
{d\tilde{r}}\left(\underset{\emph{n}-2}{\underbrace{\ldots}}\,-\frac{\delta(r,\tilde{r})}
{\gamma(\tilde{r})}\,\Psi^{(1)}\right)d\tilde{r}\\
&=&\ldots-\int \frac{d}{d\tilde{r}}\left(\frac{\partial L}{\partial
\Psi^{(\emph{n})}}\right)
\frac{1}{\gamma}\,\frac{d}{d\tilde{r}}\left(\underset{\emph{n}-3}{\underbrace{\ldots}}\,
-\frac{\delta(r,\tilde{r})}{\gamma(\tilde{r})}\,\Psi^{(1)}\right)d\tilde{r}=\\
&=&\ldots-\int \gamma\,\frac{\partial^2
L}{\partial\left(\Psi^{(\emph{n})}\right)^2}\,\Psi^{(\emph{n}+1)}
\frac{1}{\gamma}\,\frac{d}{d\tilde{r}}\left(\underset{\emph{n}-3}{\underbrace{\ldots}}\,
-\frac{\delta(r,\tilde{r})}{\gamma(\tilde{r})}\,\Psi^{(1)}\right)d\tilde{r}=\\
&& \qquad\qquad\qquad\qquad\qquad\vdots\\
&=&\ldots+(-1)^{\emph{n}-1}\int \frac{\partial^2 L}{\partial
\left(\Psi^{(\emph{n})}\right)^2}
\,\,\Psi^{(2\emph{n}-1)}\,\Psi^{(1)}\,\delta(r,\tilde{r})\,d\tilde{r}=\\
&=&\ldots+(-1)^{\emph{n}-1}\frac{\partial^2 L}{\partial
\left(\Psi^{(\emph{n})}\right)^2}
\,\,\Psi^{(2\emph{n}-1)}\,\Psi^{(1)}\,.
\end{eqnarray*}
Therefore
\begin{eqnarray*}
g^{\Phi\Phi}=\ldots-\frac{\gamma}{2\psi}(-1)^{2\emph{n}-1}\left(\frac{\partial^2
L}{\partial \left(\Psi^{(\emph{n})}\right)^2}\right)^2
\Psi^{(1)}\,\Psi^{(2\emph{n}-1)}\,\Psi^{(2\emph{n})},
\end{eqnarray*}
where the $\ldots$ stand for all other terms, not involving
$\Psi^{(2\emph{n})}$. Now, according to the aforementioned
\textbf{Requirement} we need this to be a general function, say
$F(\gamma L)$, and for this to happen the coefficient of
$\Psi^{(2\emph{n})}$ must vanish, i.e.
\begin{eqnarray*}
\frac{\partial^2 L}{\partial
\left(\Psi^{(\emph{n})}\right)^2}=0\Leftrightarrow
L=L_1\left(\Psi^{(0)},\ldots,\Psi^{(\emph{n}-1)}\right)\Psi^{(\emph{n})}+
L_2\left(\Psi^{(0)},\ldots,\Psi^{(\emph{n}-1)}\right).
\end{eqnarray*}
Now, the term in $\Phi$ corresponding to $L_1$ is, up to a surface
term, equivalent to a general term depending on
$\Psi^{(0)},\ldots,\Psi^{(\emph{n}-1)}$ only: indeed,
\begin{eqnarray*}
\Phi_1=\int
\gamma(\tilde{r})L_1\frac{1}{\gamma(\tilde{r})}\frac{d}{d\tilde{r}}\Psi^{(\emph{n}-1)}d\tilde{r},
\end{eqnarray*}
which upon subtraction of the surface term
\begin{eqnarray*}
A=\int d\tilde{r}\frac{d}{d\tilde{r}}\left(\int
d\Psi^{(\emph{n}-1)}L_1\right)
\end{eqnarray*}
produces a smooth functional with arguments up to
$\Psi^{(\emph{n}-1)}$ only. Since a surface term in $\Phi$ does not
affect the outcome of the variational derivatives
$\frac{\delta\,\Phi}{\delta\,\psi(r)}$ and
$\frac{\delta\,\Phi}{\delta\,\gamma(r)}$, we conclude that only
$L_2$ is important for the local part of $\Phi$. The entire argument
can be repeated successively for
$\emph{n}-1,\,\emph{n}-2,\ldots,\,2$; therefore all
$\Psi^{(\emph{n})}$'s are suppressed from $L$ except when
$\emph{n}=1$. The case $\emph{n}=1$ needs separate consideration
since, upon elimination of the linear in $\Psi^{(2)}$ term we are
left with a local function of $\Psi^{(1)}$, and thus the possibility
arises to meet the \textbf{Requirement} by solving a differential
equation for $L$. In more detail, if
\begin{equation}\label{Phi2}
\Phi\equiv\int
\gamma(\tilde{r})L\left(\psi,\Psi^{(1)}\right)d\tilde{r},
\end{equation}
$g^{\Phi\Phi}$ reads
\begin{eqnarray}
\nonumber g^{\Phi\Phi}&=&\frac{\gamma}{4\psi^2}\left(L-\Psi^{(1)}\,\frac{\partial
L}{\partial \Psi^{(1)}}\right)
\left[L-\Psi^{(1)}\,\frac{\partial L}{\partial
\Psi^{(1)}}-2\,\psi\left(\frac{\partial L}{\partial \psi}-
\Psi^{(1)}\,\frac{\partial^2 L}{\partial \psi\,\partial \Psi^{(1)}}\right)\right]+\\
\label{gphiphi1}&
&+\frac{\gamma}{2\psi}\left(L-\Psi^{(1)}\,\frac{\partial L}{\partial
\Psi^{(1)}}\right)\frac{\partial^2 L}{\partial
(\Psi^{(1)})^2}\,\Psi^{(2)}.
\end{eqnarray}
Through the definition
\begin{equation}\label{defH}
H\equiv L-\Psi^{(1)}\,\frac{\partial L}{\partial\Psi^{(1)}}
\end{equation}
we obtain
\begin{equation}
\frac{\partial H}{\partial \psi}=\frac{\partial L}{\partial
\psi}-\Psi^{(1)}\,\frac{\partial^2 L}{\partial \psi\,\partial
\Psi^{(1)}},\,\, \frac{\partial H}{\partial
\Psi^{(1)}}=-\Psi^{(1)}\frac{\partial^2 L}{\partial
\left(\Psi^{(1)}\right)^2}\,.\nonumber
\end{equation}
Thus (\ref{gphiphi1}) assumes the form
\begin{eqnarray*}
g^{\Phi\Phi}=\frac{\gamma}{4\psi^2}\left(H^2-2\,\psi\,H\,\frac{\partial
H}{\partial\psi}-
\frac{2\,\psi}{\Psi^{(1)}}\,H\,\frac{\partial
H}{\partial\Psi^{(1)}}\,\Psi^{(2)}\right),
\end{eqnarray*}
which upon addition, by virtue of the \textbf{Assumption}, of the
surface term
\begin{eqnarray*}
A=\frac{d}{dr}\left(\int\frac{1}{2\psi\Psi^{(1)}}\,H\,\frac{\partial
H}{\partial\Psi^{(1)}}\,d\Psi^{(1)}\right)
\end{eqnarray*}
gives
\begin{equation}\label{gphiphi2}
g^{\Phi\Phi}=\frac{\gamma}{4\psi^2}\left(H^2-2\,\psi\,H\,\frac{\partial
H}{\partial\psi}+4\psi^2\Psi^{(1)}\frac{\partial}{\partial\psi}
\int\frac{1}{2\psi\Psi^{(1)}}\,H\,\frac{\partial
H}{\partial\Psi^{(1)}}\,d\Psi^{(1)}\right).
\end{equation}
Since in the last expression we have only a multiplicative
$\gamma(r)$, it is obvious that the \textbf{Requirement}
\begin{eqnarray*}
g^{\Phi\Phi}=F(\gamma\,L)
\end{eqnarray*}
can be satisfied only by
\begin{equation}\label{gphiphi3}
g^{\Phi\Phi}=\kappa\,\gamma\,L,
\end{equation}
with $g^{\Phi\Phi}$ given by (\ref{gphiphi2}) and $\kappa$ any
constant. Upon differentiation of this equation with respect to
$\Psi^{(1)}$ we get
\begin{eqnarray*}
\frac{\partial}{\partial\psi}\int\frac{1}{2\psi\Psi^{(1)}}\,H\,\frac{\partial
H} {\partial\Psi^{(1)}}\,d\Psi^{(1)}=\kappa\,\frac{\partial
L}{\partial\Psi^{(1)}}\,.
\end{eqnarray*}
Multiplying the last expression by $\Psi^{(1)}$ and subtracting it
from (\ref{gphiphi3}) (with $g^{\Phi\Phi}$ given by
\eqref{gphiphi2}) we end up with the autonomous necessary condition
for $H(\psi,\,\Psi^{(1)})$:
\begin{eqnarray*}
H\left(\frac{1}{4\psi^2}H-\frac{1}{2\psi}\frac{\partial
H}{\partial\psi}-\kappa\right)=0,
\end{eqnarray*}
where (\ref{defH}) was also used. The above equation can be readily
integrated giving
\begin{eqnarray*}
H&=&0, \\
H&=&-\frac{4\kappa\psi^2}{3}+\sqrt{\psi}\,a(\Psi^{(1)}),
\end{eqnarray*}
where $a(\Psi^{(1)})$ is an arbitrary function of its argument.
The first possibility gives according to (\ref{defH})
$L=\lambda\,\Psi^{(1)}$ which, however, contributes to $\Phi$ a
surface term, and can thus be ignored. Inserting the second solution
into (\ref{defH}) we construct a partial differential equation for
$L$, namely
\begin{eqnarray*}
L-\Psi^{(1)}\,\frac{\partial
L}{\partial\Psi^{(1)}}=-\frac{4\kappa\psi^2}{3}+\sqrt{\psi}\,a(\Psi^{(1)}),
\end{eqnarray*}
which upon integration gives
\begin{eqnarray*}
L=-\frac{4\kappa\psi^2}{3}-\sqrt{\psi}\,\Psi^{(1)}\int\frac{a(\Psi^{(1)})}{{\Psi^{(1)}}^2}\,\,d\Psi^{(1)}+
c_1(\psi)\,\Psi^{(1)}\,.
\end{eqnarray*}
Since this form of $L$ emerged as a necessary condition, it must be
inserted (along with $H$) in (\ref{gphiphi3}). The result is that
$c_1(\psi)=0$. Thus $L$ reads
\begin{equation}\label{L}
L=-\frac{4\kappa\psi^2}{3}-\sqrt{\psi}\,\Psi^{(1)}\int\frac{a(\Psi^{(1)})}{{\Psi^{(1)}}^2}\,\,d\Psi^{(1)}\,.
\end{equation}
By assuming that the $\Psi^{(1)}$--dependent part of $L$ equals
$b(\Psi^{(1)})$, i.e.
\begin{eqnarray*}
-\Psi^{(1)}\int\frac{a(\Psi^{(1)})}{{\Psi^{(1)}}^2}\,\,d\Psi^{(1)}=b(\Psi^{(1)}),
\end{eqnarray*}
we get, upon a double differentiation with respect to
$\Psi^{(1)}$, the ordinary differential equation
\begin{eqnarray*}
-\frac{a\,'(\Psi^{(1)})}{\Psi^{(1)}}=b\,''(\Psi^{(1)})
\end{eqnarray*}
with solution
\begin{eqnarray*}
a(\Psi^{(1)})=b(\Psi^{(1)})+\kappa_1-\Psi^{(1)}\,b\,'(\Psi^{(1)}),
\end{eqnarray*}
where $\kappa_1$ is a constant. Substituting this equation into
(\ref{L}) and performing a partial integration we end up with
\begin{equation}\label{L1}
L=-\frac{4\kappa\psi^2}{3}+\kappa_1\sqrt{\psi}+\sqrt{\psi}\,b(\Psi^{(1)})\,.
\end{equation}
$\kappa$, $\kappa_1$ and $b(\Psi^{(1)})$ being completely arbitrary
and to our disposal; the two simplest choices
$\kappa=0,\,b(\Psi^{(1)})=0$ and $\kappa_1=0,\,b(\Psi^{(1)})=0$ lead
respectively to the following two basic ultra-local smooth
functionals:
\begin{equation*}
q^1=\int d\tilde{r}\gamma(\tilde{r})\sqrt{\psi(\tilde{r})}, \qquad q^2=\int
d\tilde{r}\gamma(\tilde{r})\psi(\tilde{r})^2\,.
\end{equation*}
The next simpler choice $\kappa=0,\,\kappa_1=0$ and
$b(\Psi^{(1)})$ arbitrary leads to a generic $q^3=\int
d\tilde{r}\gamma(\tilde{r})
\sqrt{\psi}\,b(\Psi^{(1)})$. However, it can
be proven that, for any choice of $b(\Psi^{(1)})$, the
corresponding renormalized induced metric
\begin{eqnarray*}
g^{AB}=G^{\alpha\beta}\frac{\delta q^A}{\delta x^\alpha}\frac{\delta
q^B}{\delta x^\beta} \qquad where \quad A,B=1,2,3
\end{eqnarray*}
is singular. The calculation of $g^{AB}$ gives:
\begin{eqnarray*}
g^{11}&=&G^{\alpha\beta}\frac{\delta q^1}{\delta
x^\alpha}\frac{\delta q^1}{\delta x^\beta}=
0\overset{Assumption}{\Longleftrightarrow}g_{ren}^{11}=0,\\
g^{12}&=&G^{\alpha\beta}\frac{\delta q^1}{\delta
x^\alpha}\frac{\delta q^2}{\delta x^\beta}=
-\frac{3}{8}\,\gamma\sqrt{\psi}\overset{Assumption}{\Longleftrightarrow}g_{ren}^{12}=-\frac{3\,q^1}{8}\,,\\
g^{22}&=&G^{\alpha\beta}\frac{\delta q^2}{\delta
x^\alpha}\frac{\delta q^2}{\delta x^\beta}=
-\frac{3}{4}\,\gamma\,\psi^2\overset{Assumption}{\Longleftrightarrow}g_{ren}^{22}=-\frac{3}{4}\,q^2,\\
g^{13}&=&G^{\alpha\beta}\frac{\delta q^1}{\delta
x^\alpha}\frac{\delta q^3}{\delta x^\beta}=
\frac{1}{4}\,\gamma\,\Psi^{(2)}b\,''=\frac{d}{dr}\left(\frac{1}{4}\,b\,'\right)\overset{Assumption}{\Longleftrightarrow}g_{ren}^{13}=0,
\end{eqnarray*}
\begin{eqnarray*}
g^{23}&=&G^{\alpha\beta}\frac{\delta q^2}{\delta
x^\alpha}\frac{\delta q^3}{\delta x^\beta}=
\frac{1}{8}\,\gamma\sqrt{\psi}\left(-3\,b+3\Psi^{(1)}b\,'+
2\,\psi\,\Psi^{(2)}\,b\,''\right)\overset{Assumption}{\Longleftrightarrow}\\
g_{ren}^{23}&=&\frac{1}{8}\int
dr\gamma\sqrt{\psi}\left(-3\,b+3\Psi^{(1)}b\,'+
2\,\psi\,\Psi^{(2)}\,b\,''\right)-\int dr\frac{d}{dr}\left(\frac{1}{4}\int d\Psi^{(1)}\psi^{3/2}\,b\,''\right)=\\
&=&-\frac{3}{8}\int dr\gamma\sqrt{\psi}\,b=-\frac{3\,q^3}{8}\,,\\
g^{33}&=&G^{\alpha\beta}\frac{\delta q^3}{\delta
x^\alpha}\frac{\delta q^3}{\delta x^\beta}=
\frac{1}{2}\,\gamma\left(b-\Psi^{(1)}\,b'\right)\Psi^{(2)}\,b\,''\overset{Assumption}{\Longleftrightarrow}\\
g_{ren}^{33}&=&\frac{1}{2}\int
dr\gamma\left(b-\Psi^{(1)}\,b'\right)\Psi^{(2)}\,b\,''-\int
dr\,\frac{d}{dr}\left[\frac{1}{2}\int
d\Psi^{(1)}\left(b-\Psi^{(1)}\,b'\right)\,b\,''\right]=0,
\end{eqnarray*}
where by $'$ we denote differentiation with respect to $\Psi^{(1)}$.
Thus the re-normalized induced metric reads
\begin{eqnarray*}
g^{AB}_{ren}(q^1,q^2,q^3)=-\frac{3}{8}
\begin{pmatrix}
\displaystyle{0} & \displaystyle{q^1} & \displaystyle{0} \cr
& \cr \displaystyle{q^1} & 2\,q^2  & q^3 \cr & \cr \displaystyle{0} & q^3 &
0 \cr
\end{pmatrix}.
\end{eqnarray*}
Effecting the transformation
$(\tilde{q}^1,\,\tilde{q}^2,\,\tilde{q}^3)=\left(q^1,\,q^2,\,f\left(
\frac{q^3}{q^1}\right)\right)$ we bring $g^{AB}_{ren}$ into a manifestly
degenerate form:
\begin{eqnarray*}
g^{AB}_{ren}(q^1,q^2)=-\frac{3}{8}
\begin{pmatrix}
\displaystyle{0} & \displaystyle{q^1} & \displaystyle{0} \cr
& \cr \displaystyle{q^1} & 2\,q^2  & 0 \cr & \cr \displaystyle{0} & 0 &
0 \cr
\end{pmatrix}.
\end{eqnarray*}
So, it seems that, as far as the ultra local part of the functionals
is concerned, the re-normalized metric is given by the upper left
$2\times2$ block of the above expression. It is interesting to
observe that the integrants of $q^1,\,q^2$ form a base in the space
spanned by $\gamma,\, \psi$. It is convenient to change these two
integrants (namely $\gamma \sqrt{\psi}$ and $\gamma\,\psi^2$) to
$\gamma$ and $\gamma\psi^2$ respectively, since the latter choice
complies with the ultra local parts of the potential
(\ref{potential}). One might wonder if this action is permitted,
namely if the new re-normalized metric, resulting from the choice of
the new functionals,
\begin{equation*}
y^1=\int\gamma(\tilde{r})d\tilde{r}, \qquad y^2=\int\gamma(\tilde{r})\psi(\tilde{r})^2d\tilde{r}
\end{equation*}
is equivalent to the previous. Interestingly enough, the answer is yes.
Indeed, following the line of thoughts leading to $g^{AB}_{ren}(q^1,q^2)$
one arrives at
\begin{eqnarray*}
g^{AB}_{ren}(y^1,y^2)=-\frac{1}{4}
\begin{pmatrix}
\displaystyle{-\frac{\left(y^1\right)^2}{y^2}} & \displaystyle{y^1} \cr
& \cr \displaystyle{y^1} & 3\,y^2 \cr
\end{pmatrix},
\end{eqnarray*}
which is related to the upper left $2\times2$ block of $g^{AB}_{ren}(q^1,q^2)$
through the transformation
\begin{equation*}
(y^1,y^2)=\left(\frac{(q^1)^{4/3}}{(q^2)^{1/3}},\,q^2\right).
\end{equation*}
Quite unexpectedly, this transformation is identical to the
transformation connecting the integrands of the two pairs of
functionals $\left(\gamma\sqrt{\psi},\gamma\psi^2\right)$,
$\left(\gamma,\gamma\psi^2\right)$. This is a strong indication that
the use of the \textbf{Assumption} preserves the geometry of the
re-normalized manifold. We thus adopt, without any loss of
generality the ultra local functionals:
\begin{equation}\label{y1,y2}
y^1=\int\gamma(\tilde{r})d\tilde{r}, \qquad y^2=\int\gamma(\tilde{r})\psi(\tilde{r})^2d\tilde{r}
\end{equation}
One might think that this preservation of the geometry of the
re-normalized manifold is due to the ultra local nature of the
integrands but, as we shall subsequently see, this state of affairs
continuous to hold even when functionals with integrands which
contain derivatives of the configuration variables are considered.

Indeed it is quite essential to have a functional that contains
first derivative of $\psi(r)$, since a term of this kind (namely
$\displaystyle{\frac{{\psi'}^2}{\gamma}}$) does appear in the
potential \eqref{potential}. Thus, it is clear that this is not the
end of our investigation for a suitable space of state vectors: the
caveat is that the argument leading to $y^1,\,y^2$ crucially depends
upon the original choice of \underline{one} initial candidate smooth
scalar functional (\ref{Phi2}). Therefore, to complete the search we
must close the circle by starting with the two already secured
smooth functionals $(y^1,\,y^2)$, and a \underline{third} of the
general form
\begin{eqnarray*}
y^3=\int d\tilde{r}\,\gamma(\tilde{r})\,L(\Psi^{(1)})
\end{eqnarray*}
(since the $\psi$ dependence has already been fixed to either $1$
or $\psi^2$). The calculation of the, related to
$y^3$, components of the induced metric $g^{AB}$ gives:
\begin{eqnarray*}
g^{13}&=&\frac{\gamma}{4\psi^2}\left(L-\Psi^{(1)}\,L'+\psi\,\Psi^{(2)}\,L''\right)
\overset{Assumption}{\Longleftrightarrow}\\
g_{ren}^{13}&=&\int
dr\frac{\gamma}{4\psi^2}\left(L-\Psi^{(1)}\,L'+\psi\,\Psi^{(2)}\,L''\right)-
\int dr\frac{d}{dr}\left(\frac{1}{4}\int d\Psi^{(1)}\,\frac{L''}{\psi}\right)=\\
&=&\frac{1}{4}\int dr\frac{\gamma}{\psi^2}\,L\overset{Assumption}{=}\frac{1}{4}\frac{\gamma}{\gamma\,\psi^2}\,\gamma
L\overset{Assumption}{=}\frac{1}{4}\frac{\int dr\gamma}{\int dr\gamma\,\psi^2}\int dr\gamma L=\frac{y^1\,y^3}{4\,y^2},\\
g^{23}&=&\frac{\gamma}{4}\left(-L+\Psi^{(1)}\,L'+\psi\,\Psi^{(2)}L''\right)
\overset{Assumption}{\Longleftrightarrow}\\
g_{ren}^{23}&=&\int
dr\frac{\gamma}{4}\left(-L+\Psi^{(1)}\,L'+\psi\,\Psi^{(2)}L''\right)
-\int dr\frac{d}{dr}\left(\frac{1}{4}\int d\Psi^{(1)}\psi\,L''\right)=\\
&=&-\frac{1}{4}\int dr\gamma\,L=-\frac{y^3}{4}\,,\\
g^{33}&=&\frac{\gamma}{4\psi^2}\left(L-\Psi^{(1)}L'\right)^2+\frac{\gamma}{2\psi}
\left(L-\Psi^{(1)}L'\right)\Psi^{(2)}L''\overset{Assumption}{\Longleftrightarrow}\\
g_{ren}^{33}&=&\int dr\left[\frac{\gamma}{4\psi^2}\left(L-\Psi^{(1)}L'\right)^2+
\frac{\gamma}{2\psi}\left(L-\Psi^{(1)}L'\right)\Psi^{(2)}L''\right]-\\
&-&\int dr\frac{d}{dr}\int
d\Psi^{(1)}\left(\frac{\left(L-\Psi^{(1)}L'\right)L''}
{2\psi}\right)\overset{Assumption}{\Longleftrightarrow}
\end{eqnarray*}
\begin{equation}\label{g33ren}
g_{ren}^{33}=\frac{\gamma}{4\psi^2}\left[\left(L-\Psi^{(1)}L'\right)^2-
\Psi^{(1)}\int\frac{d\Psi^{(1)}}{\Psi^{(1)}}\,\frac{\partial}{\partial\Psi^{(1)}}\left(L-
\Psi^{(1)}L'\right)^2\right].
\end{equation}
The expression inside the square brackets of $g_{ren}^{33}$ above,
being a generic function of $\Psi^{(1)}$, can also be considered as
a function of $L$, say $W(L(\Psi^{(1)}))$. It is thus clear that the
\textbf{Requirement} is satisfied for any $L(\Psi^{(1)})$. Let this
expression $W(L(\Psi^{(1)}))$ be parameterized as
\begin{equation}\label{g33par}
L\left(\Psi^{(1)}\right)^2-\frac{4\,{F[L\left(\Psi^{(1)}\right)]}^2}{3\,{F'[F[L\left(\Psi^{(1)}\right)]]}^2}\,.
\end{equation}
This ``peculiar'' parametrization of the arbitrariness in
$L\left(\Psi^{(1)}\right)$ has been chosen in order to facilitate
the subsequent proof that the freedom in the choice of $L$ (left by
the imposition of the \textbf{Requirement}) is a pure general
coordinate transformation (gct) of the induced re-normalized metric.

The reduced re-normalized manifold is thus parameterized by the
following three smooth scalar functionals:
\begin{equation}\label{y1,y2,y3}
y^1=\int\gamma(\tilde{r})d\tilde{r}, \qquad
y^2=\int\gamma(\tilde{r})\psi(\tilde{r})^2d\tilde{r}, \qquad
y^3=\int\gamma(\tilde{r}) L(\Psi^{(1)})d\tilde{r}.
\end{equation}

Any other functional, say $y^4=\int
d\tilde{r}\,\gamma(\tilde{r})\,K\left[\psi(\tilde{r}),\Psi^{(1)}(\tilde{r})\right]$,
can be considered as a function of $y^1,y^2,y^3$; indeed, since the
scalar functions appearing in the integrands of $y^2, y^3$ form a
base in the space spanned by $\psi,\Psi^{(1)}$, we
can express the generic $K$ in $y^4$ as
$K\left[\sqrt{\frac{\gamma\psi^2}{\gamma}},\Psi^{(1)}\right]$,
which (through the \textbf{Assumption})
gives $y^4=y^1\,K\left[\sqrt{\frac{y^2}{y^1}},L^{-1}\left(\frac{y^3}{y^1}\right)\right]$.

The geometry of this space is described by the induced re-normalized
metric
\begin{eqnarray}\label{renmet}
g^{AB}_{ren}(y^1,y^2,y^3)=-\frac{1}{4}
\begin{pmatrix}
-\displaystyle{\frac{{(y^1)}^2}{y^2}} & \displaystyle{y^1} & -\displaystyle{\frac{y^1y^3}{y^2}}
\cr & \cr \displaystyle{y^1} & 3\,y^2  &
\displaystyle{y^3} \cr & \cr -\displaystyle{\frac{y^1y^3}{y^2}} &
\displaystyle{y^3} &
-\displaystyle{\frac{{(y^3)}^2}{y^2}+\frac{4\,{(y^1)}^2F\left(\frac{y^3}{y^1}\right)^2}{3\,y^2{F'\left[F\left(\frac{y^3}{y^1}\right)\right]}^2}} \cr
\end{pmatrix}, \nonumber\\
{g_{AB}}_{ren}(y^1,y^2,y^3)=
\begin{pmatrix}
\frac{3\,y^2}{{(y^1)}^4}\left({(y^1)}^2-\frac{{(y^3)}^2{F'\left[F\left(\frac{y^3}{y^1}\right)\right]}^2}{F\left(\frac{y^3}{y^1}\right)^2}\right)& -\frac{1}{y^1} &
\frac{3\,y^2y^3{F'\left[F\left(\frac{y^3}{y^1}\right)\right]}^2}{{(y^1)}^3F\left(\frac{y^3}{y^1}\right)^2}  \cr & \cr
-\frac{1}{y^1} &
-\frac{1}{y^2} & 0 \cr & \cr
\frac{3\,y^2y^3{F'\left[F\left(\frac{y^3}{y^1}\right)\right]}^2}{{(y^1)}^3F\left(\frac{y^3}{y^1}\right)^2} & 0 &
-\frac{3\,y^2{F'\left[F\left(\frac{y^3}{y^1}\right)\right]}^2}{{(y^1)}^2F\left(\frac{y^3}{y^1}\right)^2} \cr
\end{pmatrix}.
\end{eqnarray}
Any function
$\Psi(y^1,y^2,y^3)$ on this manifold is of course annihilated by the
quantum linear constraint, i.e.
\begin{eqnarray*}
\hat{\mathcal{H}}_1\Psi(y^1,y^2,y^3)=\frac{\partial\Psi(y^1,y^2,y^3)}{\partial
y^1}\,\hat{\mathcal{H}}_1\,y^1+
\frac{\partial\Psi(y^1,y^2,y^3)}{\partial
y^2}\,\hat{\mathcal{H}}_1\,y^2+
\frac{\partial\Psi(y^1,y^2,y^3)}{\partial
y^3}\,\hat{\mathcal{H}}_1\,y^3=0
\end{eqnarray*}
since the derivatives with respect to $r$ are transparent to the
partial derivatives of $\Psi$ (which are, just like the $y^A$'s,
r-numbers).

The covariant metric \eqref{renmet} describes a three dimensional
conformally flat geometry, since the corresponding Cotton-York tensor vanishes. The
Ricci scalar is $R=\frac{3}{8\,y^2}$, indicating that the arbitrariness
in $F$ (and thus also in $L$) is a \emph{pure gauge}. The change of coordinates
\begin{equation}\label{gctransf}
(y^1,y^2,y^3)=(e^{-\frac{1}{8}(5\,Y^1+3\,Y^3)},e^{Y^1+Y^2+Y^3},
e^{-\frac{1}{8}(5\,Y^1+3\,Y^3)}F^{-1}(e^{\frac{1}{24}(-9\,Y^1+8\,Y^2-15\,Y^3)}))
\end{equation}
(where $F^{-1}$ denotes the function inverse to $F$, i.e $F^{-1}(F(x))=x$)
brings the metric to the manifestly conformally flat form:
\begin{equation}\label{conflat}
{g_{AB}}_{ren}(Y^1,Y^2,Y^3)=
\begin{pmatrix}
\displaystyle{e^{Y^1+Y^2+Y^3}} & \displaystyle{0} &
\displaystyle{0} \cr & \cr
\displaystyle{0} &
-\displaystyle{\frac{4}{3}\,e^{Y^1+Y^2+Y^3}} & 0 \cr & \cr
\displaystyle{0} & \displaystyle{0} &
-\displaystyle{e^{Y^1+Y^2+Y^3}} \cr
\end{pmatrix},
\end{equation}
in which all the $F$ dependence has indeed disappeared.

The final restriction on the form of $\Psi$ will be obtained by the
imposition of the quantum analog of the quadratic constraint
$\mathcal{H}_o$. According to the above exposition we postulate that
the quantum gravity of the geometries given by \eqref{metrics}
will be described by the following partial
differential equation (in terms of the $Y^A$'s)
\begin{equation}\label{wdw}
\hat{\mathcal{H}}_o\Psi\equiv [-\frac{1}{2}\,\Box_{c}
+V_{ren}]\,\Psi(Y^1,Y^2,Y^3)=0
\end{equation}
with
\begin{equation}\label{conf Lap}
\Box_c=\Box+\frac{d-2}{4\,(d-1)}\,R
\end{equation}
being the conformal Laplacian based on  $g_{AB\, ren}(Y^1,Y^2,Y^3)$,
$R$ the Ricci scalar, and $d$ the dimensions of $g_{AB\, ren}$. The
metric \eqref{conflat} is conformally flat with Ricci scalar
$R=\frac{3}{8}\,e^{-Y^1-Y^2-Y^3}$, and its dimension is $d=3$. The
re-normalized form of the potential (\ref{potential}) offers us the
possibility to introduce, in a dynamical way, topological effects
into our wave functional: Indeed, under our \textbf{Assumption}, the
first two terms become $-2\,\epsilon\,y^1$ and $2\,\Lambda\,y^2$,
respectively, while the last, being a total derivative, becomes $
A_T \equiv 4\,\frac{\psi\,\psi^\prime}{\gamma}\mid^\beta_\alpha
\,(if\,\alpha<r<\beta) $. In the spirit previously explained we
should drop this term, however one could also keep it. The
re-normalized form of the remaining, third, term of the potential
can be obtained as follows
\begin{eqnarray*}
y^3&=&\gamma\,L(\Psi^{(1)})\Leftrightarrow L(\Psi^{(1)})=\frac{y^3}{\gamma}
\overset{Assumption}{\Longleftrightarrow}L(\Psi^{(1)})=\frac{y^3}{y^1}\Leftrightarrow
\Psi^{(1)}=L^{-1}\left(\frac{y^3}{y^1}\right),
\end{eqnarray*}
thus finally
\begin{equation*}
\frac{\psi'}{\gamma}=L^{-1}\left(\frac{y^3}{y^1}\right)
\end{equation*}
and the third term becomes
$-2\,y^1\left[L^{-1}\left(\frac{y^3}{y^1}\right)\right]^2$. Finally,
effecting the transformation \eqref{gctransf} the form of the
re-normalized potential is
\begin{eqnarray}
V_{ren}&=&-2\,\epsilon\,e^{-\frac{1}{8}(5\,Y^1+3\,Y^3)}
-2\,e^{-\frac{1}{8}(5\,Y^1+3\,Y^3)}\left[L^{-1}\left(F^{-1}(e^{\frac{1}{24}(-9\,Y^1+8\,Y^2-15\,Y^3)})\right)\right]^2+\nonumber\\
&&2\,\Lambda\,e^{Y^1+Y^2+Y^3}+A_T \label{conpot}
\end{eqnarray}
and the Wheeler-deWitt equation is given as
\begin{eqnarray*}
&&-2\,\epsilon\,e^{\frac{1}{8}(3\,Y^1+5\,Y^3)+Y^2}\,\Psi(Y^1,Y^2,Y^3)+2\,\Lambda\,e^{2(Y^1+Y^2+Y^3)}\,\Psi(Y^1,Y^2,Y^3)-\\
&&2\,e^{\frac{1}{8}(3\,Y^1+5\,Y^3)+Y^2}\left[L^{-1}\left(F^{-1}(e^{\frac{1}{24}(-9\,Y^1+8\,Y^2-
15\,Y^3)})\right)\right]^2\Psi(Y^1,Y^2,Y^3)+\\
&&A_T\,e^{Y^1+Y^2+Y^3}\,\Psi(Y^1,Y^2,Y^3)-\frac{3}{128}\Psi(Y^1,Y^2,Y^3)-
\frac{1}{4}\frac{\partial\Psi(Y^1,Y^2,Y^3)}{\partial Y^1}+\\
&&\frac{3}{16}\frac{\partial\Psi(Y^1,Y^2,Y^3)}{\partial Y^2}+
\frac{1}{4}\frac{\partial\Psi(Y^1,Y^2,Y^3)}{\partial Y^3}-\frac{1}{2}\frac{\partial^2\Psi(Y^1,Y^2,Y^3)}{\partial\left(Y^1\right)^2}+
\frac{3}{8}\frac{\partial^2\Psi(Y^1,Y^2,Y^3)}{\partial\left(Y^2\right)^2}+\\
&&\frac{1}{2}\frac{\partial^2\Psi(Y^1,Y^2,Y^3)}{\partial\left(Y^3\right)^2}=0.
\end{eqnarray*}
Since $F^{-1}$ is an arbitrary function of its arguments, we may
contemplate the choice:
\begin{equation}\label{F}
F^{-1}\left(e^{\frac{1}{24}(-9\,Y^1+8\,Y^2-15\,Y^3)}\right)=L\left(\sqrt{e^{\frac{1}{24}(-9\,Y^1+8\,Y^2-15\,Y^3)}-
\epsilon}\,\right).
\end{equation}
Of course there is a question of existence for such a choice: since
$F$ which appears in \eqref{g33par} is a convenient parametrization
of \eqref{g33ren}, any demand that $F$ has a specified form (much
more in terms of $L$) constitutes an implicit restriction on the form
of $L$ itself. Subsequently, the existence of such an $L$ must be proven.
Indeed, in the Appendix $A$ it is shown that an appropriate $L$ exists,
and its form is given by \eqref{L(w)}:
\begin{equation*}
L(\Psi^{(1)})=m+\int{\frac{(\Psi^{(1)})^{3/2}}{{((\Psi^{(1)})}^2-\epsilon)^{13/16}}\,
e^{k-\frac{3\,\epsilon}{16((\Psi^{(1)})^2-\epsilon)}}\,d\Psi^{(1)}}\,\,\,
where \,\,\,\, c_1\,m+c_2+c_3\,e^k=0.
\end{equation*}
This choice for $F$
reduces the Wheeler-deWitt equation to the final separable form
\begin{eqnarray}\label{WdW}
&&2\,\Lambda\,e^{2(Y^1+Y^2+Y^3)}\Psi(Y^1,Y^2,Y^3)-2\,e^{\frac{4}{3}\,Y^2}\Psi(Y^1,Y^2,Y^3)+
A_T\,e^{Y^1+Y^2+Y^3}\Psi(Y^1,Y^2,Y^3)-\nonumber\\
&&\frac{3}{128}\Psi(Y^1,Y^2,Y^3)-\frac{1}{4}\frac{\partial\Psi(Y^1,Y^2,Y^3)}{\partial Y^1}+
\frac{3}{16}\frac{\partial\Psi(Y^1,Y^2,Y^3)}{\partial Y^2}+
\frac{1}{4}\frac{\partial\Psi(Y^1,Y^2,Y^3)}{\partial Y^3}-\nonumber\\
&&\frac{1}{2}\frac{\partial^2\Psi(Y^1,Y^2,Y^3)}{\partial\left(Y^1\right)^2}+
\frac{3}{8}\frac{\partial^2\Psi(Y^1,Y^2,Y^3)}{\partial\left(Y^2\right)^2}+
\frac{1}{2}\frac{\partial^2\Psi(Y^1,Y^2,Y^3)}{\partial\left(Y^3\right)^2}=0.
\end{eqnarray}
This equation is separable for $\Lambda=0$ and $A_T=0$. In this case
it can readily be solved: assuming
$\Psi(Y^1,Y^2,Y^3)=\Psi^1(Y^1)\,\Psi^2(Y^2)\,\Psi^3(Y^3)$ and
dividing (\ref{WdW}) by $\Psi$ we get the three ordinary
differential equations: \bsub
\begin{eqnarray*}
&&\frac{1}{4\,\Psi^1(Y^1)}\frac{d\Psi^1(Y^1)}{dY^1}+
\frac{1}{2\,\Psi^1(Y^1)}\frac{d\,^2 \Psi^1(Y^1)}{d\left(Y^1\right)^2}=m, \\
&&\frac{3}{16\,\Psi^2(Y^2)}\frac{d\Psi^2(Y^2)}{dY^2}+
\frac{3}{8\,\Psi^2(Y^2)}\frac{d\,^2 \Psi^2(Y^2)}{d\left(Y^2\right)^2}-
2\,e^{\frac{4}{3}\,Y^2}=n,\\
&&\frac{1}{4\,\Psi^3(Y^3)}\frac{d\Psi^3(Y^3)}{dY^3}+
\frac{1}{2\,\Psi^3(Y^3)}\frac{d\,^2 \Psi^3(Y^3)}{d\left(Y^3\right)^2}-
\frac{3}{128}=m-n,
\end{eqnarray*}
\esub where $m$ and $n$ are separation constants. Their solutions
are: \bsub
\begin{eqnarray*}
\Psi^1(Y^1)&=&c_1\,e^{\frac{1}{4}\left(-1-\sqrt{1+32\,m}\right)Y^1}+
c_2\,e^{\frac{1}{4}\left(-1+\sqrt{1+32\,m}\right)Y^1}\,,\\
\Psi^2(Y^2)&=&c_3\,e^{-Y^2/4}\,I_{-\frac{\sqrt{3}}{8}\sqrt{3+128\,n}}\left(2\sqrt{3}\,e^{2\,Y^2/3}\right)
+c_4\,e^{-Y^2/4}\,I_{\frac{\sqrt{3}}{8}\sqrt{3+128\,n}}\left(2\sqrt{3}\,e^{2\,Y^2/3}\right)\\
\Psi^3(Y^3)&=&c_5\,e^{\frac{1}{8}\left(-2-\sqrt{7+128\,m-128\,n}\right)Y^3}+
c_6\,e^{\frac{1}{8}\left(-2+\sqrt{7+128\,m-128\,n}\right)Y^3}\,,
\end{eqnarray*}\esub
where $I_{\pm
\frac{\sqrt{3}}{8}\sqrt{3+128\,n}}\left(2\sqrt{3}\,e^{2\,Y^2/3}\right)$
are modified Bessel functions of the first kind and non-integer order.

\section{Discussion}
We have considered the canonical analysis and subsequent
quantization of the (3+1)-dimensional action of pure gravity plus a
cosmological constant term, under the assumption of the existence of
two-dimensional (spacelike) surfaces of maximal symmetry. At the
classical level, the application of the Dirac algorithm results in
one linear and one quadratic first class constraints. The linear
constraint is shown to correspond to arbitrary changes of the radial
coordinate. The quadratic constraint is the generator of the time
evolution. Adopting the Schr\"{o}dinger picture for the quantum
momentum operators, we are led to choose as our initial collection
of state vectors all smooth (integrals over the radial coordinate
$r$) functionals, in order to avoid an ill-defined action of these
operators. The quantum linear constraint entails a reduction of this
collection to all smooth scalar functionals. At this stage the need
emerges to somehow obtain an induced metric on the so far
``physical'' states, which is composed out of these states. This
leads us to firstly adopt a particular (formal) re-normalization
prescription and secondly impose the \textbf{Requirement}. As a
result, the final collection of state vectors is reduced to the
three (essentially unique) smooth scalar functionals
($y^1,y^2,y^3$). The quantum analogue of the kinetic part of the
quadratic constraint is then realized as the conformal
Laplace-Beltrami operator based on the induced re-normalized metric.
After the interpretation (through the \textbf{Assumption}) of the
potential part of $H_o$ a Wheeler-deWitt equation emerges. In order
to analytically solve this equation we exploit the freedom in the
choice of $L$ appearing in $y^3$, which is left by the imposition of
\textbf{Requirement} and which is shown to be a \emph{pure} general
coordinate transformation on the re-normalized manifold. Effecting
an appropriate change of variables the metric is put in conformally
flat form. Then, the aforementioned freedom is used to make the
equation separable.

Generally (and somewhat loosely) speaking, our goal is, at a first
stage, to assign a unique number between 0 and 1 to each and every
geometry of the families considered, in a way that is independent of
the coordinate system used to represent the metric. Of course, at
the present status of things we cannot do this, since the following
two problems remain to be solved: i) render finite the three smooth
functionals and ii) select an appropriate inner product.

The first will need a final regularization of $y^1,y^2,y^3$, but
most probably, the detailed way to do this will depend upon the
particular geometry under consideration.

For the second, a natural choice would be the determinant of the
induced re-normalized metric, although the problem with the positive
definiteness may dictate another choice.

Finally, we would like to comment upon the relation of the results
here obtained to the results presented in our previous work
\cite{tchris_2+1}. There, the 2+1 action of pure gravity plus a
cosmological constant term was quantized in a similar manner, under
the assumption of existence of \emph{one} space-like Killing vector
field. We would like to point out the quite interesting fact that,
although the systems considered are different, the resulting
re-normalized manifolds, their geometry and the corresponding
Wheeler--deWitt equations are strikingly similar. To our view, this
constitutes a very strong indication that the imposition of the {\bf
Assumption} and the {\bf Requirement} is not simply an elegant way
to reduce the number of spatial derivatives of the configuration
fields involved in the scalar functionals, but is rather a tool for
unraveling the underlying geometrical structure of Quantum Gravity
(in the approximation considered, of course).

 \section*{Acknowledgments}
One of the authors (G. O. Papadopoulos) is a Killam Postdoctoral
Fellow and acknowledges the relevant support from the Killam
Foundation.

\newpage

\appendix

\section{Appendix: Existence of $L$}

In this Appendix we show that an $L(\Psi^{(1)})$ exists for which
\eqref{g33ren} is equal to \eqref{g33par} for the particular choice
of $F[L(\Psi^{(1)})]$ given by \eqref{F}. To begin with let us
change coordinates from $\Psi^{(1)}$ to
$\omega\equiv\sqrt{{\Psi^{(1)}}^2-\epsilon}$. The term inside the
square brackets in the r.h.s of \eqref{g33ren} becomes
\begin{equation*}
\left(L(\omega)-\frac{\epsilon+\omega^2}{\omega}L'(\omega)\right)^2-2\sqrt{\epsilon+
\omega^2}\int{\frac{\left(L(\omega)-\frac{\epsilon+\omega^2}{\omega}L'(\omega)\right)\left(L(\omega)-
\frac{\epsilon+\omega^2}{\omega}L'(\omega)\right)'}{\sqrt{\epsilon+\omega^2}}\,d\omega},
\end{equation*}
where the prime now denotes differentiation with respect to the
variable $\omega$. On the other hand \eqref{g33par}, through the
choice $F=L^{-1}(\omega)$ becomes
\begin{equation*}
L(\omega)^2-\frac{4\,\omega^2}{3}{L'(\omega)}^2.
\end{equation*}
We thus have to prove the existence of an $L(\omega)$ which secures
the equality between the above two expressions, namely that
\begin{eqnarray}
I &\equiv&
\left(L(\omega)-\frac{\epsilon+\omega^2}{\omega}L'(\omega)\right)^2-2\sqrt{\epsilon+
\omega^2}\int{\frac{\left(L(\omega)-\frac{\epsilon+\omega^2}{\omega}L'(\omega)\right)\left(L(\omega)-
\frac{\epsilon+\omega^2}{\omega}L'(\omega)\right)'}{\sqrt{\epsilon+\omega^2}}\,d\omega} \nonumber\\
&-&L(\omega)^2+\frac{4\,\omega^2}{3}{L'(\omega)}^2=0. \label{A1}
\end{eqnarray}
Let us assume that \eqref{A1} holds. Then, the expression
$-3\,\omega^2\left(I-\frac{\epsilon+\omega^2}{\omega}\frac{\partial
I}{\partial\omega}\right)$ must also vanish, which leads to
\begin{equation}\label{A2}
-3(\epsilon+\omega^2)^2L'(\omega)^2+4\,\omega^2(2\,\epsilon+
\omega^2)L'(\omega)^2+8\,\omega^3(\epsilon+\omega^2)L'(\omega)L''(\omega)=0.
\end{equation}
The case $L(\omega)=const.$ does not concern us since it corresponds to
the functional $y^1$. Therefore, dividing \eqref{A2} by
$L'(\omega)^2$ and defining $a(\omega)\equiv L'(\omega)/L(\omega)$
we obtain
\begin{equation}\label{A3}
8\,\omega^3(\epsilon+\omega^2)\,a'(\omega)+\omega^4+2\,\epsilon\,\omega^2-3\,\epsilon^2=0,
\end{equation}
which is readily integrated, giving
\begin{equation*}
a(\omega)=k+\frac{1}{8}\left(-\frac{3\,\omega}{2\,\omega^2}-5\ln{\omega}+2\ln{(\epsilon+\omega^2)}\right)
\end{equation*}
and thus
\begin{equation}\label{L(w)}
L(\omega)=m+\int{\frac{(\epsilon+\omega^2)^{1/4}}{\omega^{5/8}}\,e^{k-\frac{3\,\epsilon}{16\,\omega^2}}\,d\omega}.
\end{equation}
This expression for $L(\omega)$ emerged as an integrability
condition for the integro-differential equation \eqref{A1}. It is
therefore necessary to insert \eqref{L(w)} into this equation. The
result is the following expression for
$A\equiv-4(\epsilon+\omega^2)^{-1/2}I$:
\begin{eqnarray*}
A&=&-\frac{32\,\omega^{3/4}}{3}\,e^{2\,k-\frac{3\,\epsilon}{8\,\omega^2}}+
\frac{8\,(\epsilon+\omega^2)^{3/4}}{\omega^{13/8}}\,e^{k-\frac{3\,\epsilon}{16\,\omega^2}}\left(m+
\int{\frac{(\epsilon+\omega^2)^{1/4}}{\omega^{5/8}}\,e^{k-\frac{3\,\epsilon}{16\,\omega^2}}\,d\omega}\right)-\\
&&\int{\left[\frac{(3\,\epsilon^2-10\,\epsilon\,\omega^2-\omega^4)}{\omega^{37/8}(\epsilon+\omega^2)^{1/4}}\,e^{k-
\frac{3\,\epsilon}{16\,\omega^2}}\left(m+\int{\frac{(\epsilon+\omega^2)^{1/4}}{\omega^{5/8}}\,e^{k-
\frac{3\,\epsilon}{16\,\omega^2}}\,d\omega}\right)\right]d\omega}.
\end{eqnarray*}
Surprisingly enough the above expression is $\omega$-independent,
i.e $\partial A/\partial\omega=0$. Therefore
\begin{equation}\label{defB}
B\equiv A-F(k,m)=0.
\end{equation}
We have now to prove that there is a choice of the constants $k,m$
for which $F=0$, so that $A=0\Rightarrow I=0$. Our strategy is to
confine, through integrability conditions for \eqref{defB}, as much
as possible the form of $F(k,m)$. As a first step, we must get rid of the double
integral (in the variable $\omega$) appearing in $A$. To do this we
differentiate $B$ with respect to $k$ and solve the resulting expression
for the aforementioned double integral. By inserting the outcome of
this operation into \eqref{defB} we get
\begin{equation*}
B=\frac{\partial F}{\partial k}-2\,F+\frac{8\,m\,(\epsilon+\omega^2)^{3/4}}{\omega^{13/8}}\,e^{k-
\frac{3\,\epsilon}{16\,\omega^2}}-m\int{\frac{(3\,\epsilon^2-10\,\epsilon\,\omega^2-\omega^4)}{\omega^{37/8}(\epsilon+\omega^2)^{1/4}}\,e^{k-
\frac{3\,\epsilon}{16\,\omega^2}}\,d\omega}=0.
\end{equation*}
By differentiating this new form $B$ with respect to $k$, and repeating
the procedure described just above, we can eliminate the remaining integral
that appears in $B$ (and in fact all the $\omega$-dependence). Thus, we end up with
\begin{equation*}
B=2\,F-3\,\frac{\partial F}{\partial k}+\frac{\partial^2F}{\partial k^2}=0,
\end{equation*}
which has the following solutions:
\begin{equation}\label{F(k,m)}
F(k,m)=\lambda_1(m)\,e^k+\lambda_2(m)\,e^{2\,k}.
\end{equation}
Inserting \eqref{F(k,m)} into \eqref{defB} and differentiating with respect
to $m$ we get
\begin{equation*}
\frac{8\,(\epsilon+\omega^2)^{3/4}}{\omega^{13/8}}\,e^{k-
\frac{3\,\epsilon}{16\,\omega^2}}-\int{\frac{(3\,\epsilon^2-10\,\epsilon\,\omega^2-\omega^4)}{\omega^{37/8}(\epsilon+\omega^2)^{1/4}}\,e^{k-
\frac{3\,\epsilon}{16\,\omega^2}}\,d\omega}-{\lambda_1}'(m)\,e^k-{\lambda_2}'(m)\,e^{2\,k}=0,
\end{equation*}
which by differentiation with respect to $k$ gives
\begin{equation*}
\frac{8\,(\epsilon+\omega^2)^{3/4}}{\omega^{13/8}}\,e^{k-
\frac{3\,\epsilon}{16\,\omega^2}}-\int{\frac{(3\,\epsilon^2-10\,\epsilon\,\omega^2-\omega^4)}{\omega^{37/8}(\epsilon+\omega^2)^{1/4}}\,e^{k-
\frac{3\,\epsilon}{16\,\omega^2}}\,d\omega}-{\lambda_1}'(m)\,e^k-2\,{\lambda_2}'(m)\,e^{2\,k}=0.
\end{equation*}
Subtracting these last two equations we have
\begin{equation*}
{\lambda_2}'(m)=0\Rightarrow\lambda_2(m)=c_3.
\end{equation*}
If we insert this result together with \eqref{F(k,m)} in \eqref{defB} and we
double differentiate with respect to $m$ we will get
\begin{equation*}
{\lambda_1}''(m)=0\Rightarrow\lambda_1(m)=c_1\,m+c_2.
\end{equation*}
So, \eqref{F(k,m)} becomes
\begin{equation*}
F(k,m)=c_1\,m\,e^k+c_2\,e^k+c_3\,e^{2\,k}.
\end{equation*}
So, we finally conclude that the choice $m=-\frac{c_2+c_3\,e^k}{c_1}$
satisfies \eqref{A1}.

\end{document}